\documentclass[a4paper]{ifacconf}

\usepackage{amsmath,amssymb,amsfonts,mathtools,bm}
\usepackage{graphicx}
\usepackage{textcomp}  

\usepackage{enumerate}
\usepackage[super]{nth}
\usepackage{array}
\usepackage{booktabs}
\usepackage[para]{threeparttable}
\newlength{\abovecaptionskip}  
\usepackage{tabularx}
\usepackage[acronym]{glossaries}

\counterwithin*{section}{part}

\usepackage[capitalize,noabbrev]{cleveref}

\usepackage{siunitx}

\usepackage[normalem]{ulem}
\usepackage{natbib}        

\crefname{equation}{}{}
\crefname{problem}{Problem}{Problems}

\DeclareMathOperator{\sect}{sect}
\DeclareMathOperator{\trace}{trace}

\newcommand{\varStructK}{\bm{\mathcal{K}}}

\newacronym{are}{ARE}{algebraic Riccati equation}
\newacronym{desy}{DESY}{Deutsches Elektronen-Synchrotron}
\newacronym{euxfel}{European X-Ray Free-Electron Laser}{European XFEL}
\newacronym[longplural={frequency domain inequalities}]{fdi}{FDI}{frequency domain inequality}
\newacronym{hpc}{HPC}{high-performance computing}
\newacronym{iqc}{IQC}{integral quadratic constraint}
\newacronym{kyp}{KYP}{Kalman–Yakubovich–Popov}
\newacronym{lbsync}{LbSync}{laser-based optical synchronization}
\newacronym{lfr}{LFR}{linear fractional representation}
\newacronym{lft}{LFT}{linear fractional transformation}
\newacronym[longplural={linear matrix inequalities}]{lmi}{LMI}{linear matrix inequality}
\newacronym{lti}{LTI}{linear time-invariant}
\newacronym{pi}{PI}{proportional-integral}
\newacronym{pll}{PLL}{phase-locked loop}
\newacronym{rf}{RF}{radio-frequency}
\newacronym{rms}{RMS}{root mean square}
\newacronym{sdp}{SDP}{semi-definite program}
\newacronym{si}{SI}{sparsity invariance}
\newacronym{snr}{SNR}{signal-to-noise ratio}
\newacronym{xfel}{XFEL}{X-ray free-electron laser}

\glsunset{euxfel}
\glsunset{lti}
\glsunset{rms}
\glsunset{pi}
\glsunset{rf}

\usepackage[absolute]{textpos}
\textblockorigin{1.5cm}{\paperheight-1cm}
\begin{document}
\begin{frontmatter}
	\title{\Large \bf
		Structured IQC Synthesis of Robust \(\mathcal{H}_2\)~Controllers in the Frequency Domain\thanksref{footnoteinfo}}

	\thanks[footnoteinfo]{The authors acknowledge support from Deutsches Elektronen-Synchrotron DESY Hamburg, Germany, a member of the Helmholtz Association HGF. \copyright\ All figures and pictures under a CC BY 4.0 license.}

	\author[First,Second]{Maximilian Sch\"utte}
	\author[First,Second]{Annika Eichler}
	\author[Second]{Herbert Werner}

	\address[First]{Deutsches Elektronen-Synchrotron DESY, Germany \\ (e-mail: maximilian.schuette@desy.de).}
	\address[Second]{Institute of Control Systems, Technical University Hamburg, Germany \\ (e-mail: h.werner@tuhh.de)}

	\begin{abstract}
		The problem of robust controller synthesis for plants affected by structured uncertainty, captured by \acrlongpl{iqc}, is discussed. The solution is optimized towards a worst-case white noise rejection specification, which is a generalization of the standard \(\mathcal{H}_2\)-norm to the robust setting including possibly non-\gls{lti} uncertainty. Arbitrary structural constraints can be imposed on the control solution, making this method suitable for distributed systems. The nonsmooth optimization algorithm used to solve the robust synthesis problem operates directly in the frequency domain, eliminating scalability issues for complex systems and providing local optimality certificates. The method is evaluated using a literature example and a real-world system using a novel implementation of a robust \(\mathcal{H}_2\)-performance bound.
	\end{abstract}

	\begin{keyword}
		Distributed robust controller synthesis, Numerical methods for optimal control, Controller constraints and structure, Complex systems, H2-performance, IQCs
	\end{keyword}
\end{frontmatter}
\thispagestyle{plain}
\begin{textblock*}{\paperwidth-1cm}(0mm,0mm)
	\footnotesize Copyright © 2023 The Authors. This work has been accepted to IFAC for publication under a Creative Commons Licence CC-BY-NC-ND. \\
	Peer review under responsibility of International Federation of Automatic Control. \\
	10.1016/j.ifacol.2023.10.1055
\end{textblock*}

\section{Introduction} \label{sec:introduction}
In recent years, modern control has produced flexible and computationally efficient tools to analyze and synthesize controllers for plants affected by uncertainty from a variety of classes, from simple static unknown parameters to uncertain time delays and entirely non-linear dynamics. Current state-of-the-art methods include the use of \glspl{iqc} in combination with the \gls{kyp} lemma, see e.g.\ \cite{veenman2016}, which can be used not only to guarantee robust stability but at the same time also various performance bounds for uncertain closed-loop systems with fixed controllers. This and other \gls{lmi} based methods, however quickly yield non-convex optimization problems in the synthesis case, possibly hardened by additional constraints on the controller structure such as output-feedback or decentralized control, which are often imposed by increasingly complex control applications. Solving these problems is either achieved by exploiting assumptions on the uncertainty, the interconnection structure, \cite{eichler2014}, using iterative methods, \cite{veenman2014}, or convex approximations, \cite{furieri2020b}.

All these methods have downsides such as diminished applicability, increased conservatism or lack of optimality certificates, which is why in \cite{cavalcanti2020} a different approach is pursued. In this recent work, the authors use nonsmooth optimization techniques, see \cite{apkarian2006}, to solve the \gls{iqc} robust synthesis problem directly in the frequency domain, thus yielding the full flexibility of the \gls{iqc} framework along with local certificates of optimality and the ability to dictate the controller structure. The authors show that their method is directly compatible with existing nonsmooth optimization tools such as MATLAB's \texttt{SYSTUNE}.

Among the commonly used performance specifications, \(\mathcal{H}_2\)-performance is of particular interest to practitioners due to its inherent association with the closed-loop system's ability to mitigate the influence of white Gaussian noise. An apparent shortcoming of the framework of \cite{cavalcanti2020} is therefore that it is not possible to formulate a robust \(\mathcal{H}_2\)-performance problem with the current capabilities of \texttt{SYSTUNE}. After recapitulating the nonsmooth \gls{iqc} optimization method in \cref{sec:iqc}, the main contributions of this work hence lie in \cref{sec:h2_mult}, where compatibility of the \gls{iqc} framework and a robust \(\mathcal{H}_2\)-performance test proposed by \cite{paganini1999} is shown, and in \cref{sec:implementation}, where an efficient implementation of said test is outlined. Finally, in \cref{sec:results}, we evaluate the method using an artificial example that highlights the performance benefit of the local optimality certificate, and a demanding real-world structured controller synthesis problem.

\vspace*{-5mm}
\subsubsection{Notation.}
Let \(\bar{\sigma} (X) \) denote the largest singular value of \(X\). The bilinear sector transformation is given by \(\sect(X) = (I - X)(I + X)^{-1}\). The space of real rational proper transfer matrices without poles on the extended imaginary axis is denoted \( \mathcal{RL}_\infty^{\bullet \times \bullet} \). Its subspace \(\mathcal{RH}_\infty^{\bullet \times \bullet}\) contains those elements without poles in the closed right-half plane. \(\mathcal{S}_\infty^{\bullet \times \bullet}\) is the subset of hermitian valued members of \(\mathcal{RL}_\infty^{\bullet \times \bullet}\). Unless at risk of ambiguity, we will omit explicitly stating the dimensionality of spaces, operators and signals.

\section{IQC Synthesis via Nonsmooth Optimization} \label{sec:iqc}
\begin{figure}
	\centering
	\includegraphics[scale=0.8]{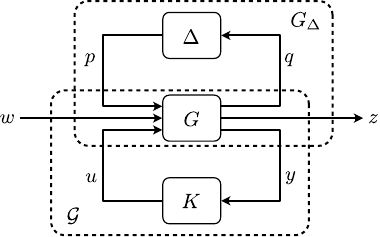}
	\caption{\Acrlong{lfr} of an uncertain system \(\mathcal{G}(\Delta)\).}
	\label{fig:lfr}
\end{figure}

We consider an uncertain distributed system consisting of arbitrarily interconnected subsystems in standard \gls{lfr} as depicted in \cref{fig:lfr} and described by
\begin{equation} \label{eq:sys}
	\begin{pmatrix} q \\ z \\ y \end{pmatrix} = G \begin{pmatrix}
		p \\ w \\ u
	\end{pmatrix},\; p = \Delta q ,\; u = K y \text{.}
\end{equation}
The uncertainty channel \(p \rightarrow q\) is closed by the bounded causal operator \(\Delta \in \bm{\Delta}\), where the compact set \(\bm{\Delta}\) reflects the uncertainty in the system model and we require the common assumption that for \(\rho \in [0,1]\) also \(\rho\Delta \in \bm{\Delta}\). The uncertain open-loop system \(G_\Delta\) is given by the upper \acrlong{lft} of its nominal \gls{lti} part \(G\) and \(\Delta\), \(G_\Delta = \mathcal{F}_u(G, \Delta)\). Similarly, we have \( \mathcal{G} = \mathcal{F}_l(G, K) \), where \(K \in \varStructK\) is a sought-after distributed output-feedback controller, whose required structure is encoded in the set of admissible dynamic controllers \(\mathcal{\varStructK}\), that closes the control channel \(y \rightarrow u\) and minimizes a worst-case performance criterion such as the \(\mathcal{H}_2\) or \(\mathcal{H}_\infty\)-norm on the performance channel \(w \rightarrow z\).

The \gls{iqc} framework, \cite{megretski1997}, provides a powerful tool for analyzing such interconnections for a broad range of uncertainties and finding suitable stabilizing and performance-optimizing controllers. We refer the reader to \cite{veenman2016} for an in-depth treatment, but will in the following also introduce relevant parts of the theory in accordance with \cite{cavalcanti2020} required for the implementation and numerical results presented in subsequent sections.

We start by introducing multipliers \(\Pi, \Pi_p \in \bm{\Pi}\) for the robustness and performance tests, respectively. The set of dynamic \gls{iqc} multipliers \(\bm{\Pi}\) imposes a virtually non-restrictive structure according to
\begin{equation} \label{eq:mult_struct}
	\bm{\Pi} = \left\{ \Pi = \begin{pmatrix}
		\Sigma^H \Sigma & \phi_o^H          \\
		\phi_o          & \Psi + \Psi^H
	\end{pmatrix} \right\} \text{,}
\end{equation}
where \(\Sigma, \phi_o, \Psi \in \mathcal{RL}_\infty\). Explicitly, we will mean these terms to constitute the so-called augmented multiplier \(\Pi_a = \mathcal{C} (\Pi, \Pi_p)\), where the composition \(\mathcal{C} \colon \bm{\Pi} \times \dots \times \bm{\Pi} \to \bm{\Pi}\) indicates block diagonal concatenation of the respective terms of its arguments, i.e.\ for
\begin{gather*}
	\Pi_i = \begin{pmatrix}
		\Sigma_i^H \Sigma_i & \phi_{o,i}^H \\
		\phi_{o,i} & \Psi_i + \Psi_i^H
	\end{pmatrix}, \quad i \in \left\{1,2\right\}
\end{gather*}
we get
\begin{multline*}
	\mathcal{C}(\Pi_1, \Pi_2) \\
	= \begin{pmatrix}
		\begin{bmatrix}
			\Sigma_1 & \\ & \Sigma_2
		\end{bmatrix}^H \begin{bmatrix}
		\Sigma_1 & \\ & \Sigma_2
	\end{bmatrix} & \begin{bmatrix}
		\phi_{o,1} & \\ & \phi_{o,2}
		\end{bmatrix}^H \\
	\begin{bmatrix}
		\phi_{o,1} & \\ & \phi_{o,2}
	\end{bmatrix} & \begin{bmatrix}
		\Psi_1 & \\ & \Psi_2
		\end{bmatrix} + \begin{bmatrix}
		\Psi_1 & \\ & \Psi_2
		\end{bmatrix}^H
	\end{pmatrix} \text{.}
\end{multline*}
Note that \(\Pi\) and \(\Pi_p\) may themselves be constructed in the same fashion from smaller multipliers to match a block diagonal structure of \(\Delta\) or to impose different performance criteria on contingent blocks of \(z\), respectively. With this notation, we state the following robust analysis result adapted from \cite{cavalcanti2020}:
\begin{cor} \label{c:iqc}
	Consider interconnection \cref{eq:sys} for a fixed \(K \in \varStructK\) s.t.\ \(\mathcal{G} \in \mathcal{RH}_\infty \) and let \(\Delta \in \bm{\Delta}\) be a bounded causal operator. Assume that:
	\begin{enumerate}[i)]
		\item for all \(\Delta \in \bm{\Delta}\), the interconnection \cref{eq:sys} is well-posed;
		\item for all \(\Delta \in \bm{\Delta}\), the following \gls{iqc} defined by some \(\Pi \in \bm{\Pi}\) is satisfied \\
		      \begin{equation} \label{eq:iqc}
			      \int_{-\infty}^{\infty} \begin{pmatrix}
				      \hat{q} (i \omega) \\
				      \hat{p} (i \omega)
			      \end{pmatrix}^H \Pi (i \omega) \begin{pmatrix}
				      \hat{q} (i \omega) \\
				      \hat{p} (i \omega)
			      \end{pmatrix} \,\mathrm{d}\omega \geq 0 \text{;}
		      \end{equation}
		\item the condition
		      \begin{equation} \label{eq:singular_valuecond}
			      \bar{\sigma} \left(\sect \left( - \begin{bmatrix}
						      \phi_o (i \omega) \mathcal{G} (i \omega) + \Psi (i \omega) & 0               \\
						      \Sigma (i \omega) \mathcal{G} (i \omega)                   & - \frac{1}{2} I
					      \end{bmatrix} \right) \right) < 1
		      \end{equation}
		      is satisfied for some \(\Pi_p \in \bm{\Pi}\) and for all \(\omega \in \mathbb{R}\), where
		      \begin{equation*}
		      	\begin{pmatrix}
		      		\Sigma^H \Sigma & \phi_o^H      \\
		      		\phi_o          & \Psi + \Psi^H
		      	\end{pmatrix} = \mathcal{C}(\Pi,\Pi_p) \text{.}
		      \end{equation*}
	\end{enumerate}
	Then, the interconnection \cref{eq:sys} is robustly stable and robust performance on the channel \(w \rightarrow z\) w.r.t.\ the criterion
	\begin{equation}
		\int_{-\infty}^{\infty} \begin{pmatrix}
			\hat{z} (i \omega) \\
			\hat{w} (i \omega)
		\end{pmatrix}^H \Pi_p \begin{pmatrix}
			\hat{z} (i \omega) \\
			\hat{w} (i \omega)
		\end{pmatrix} \,\mathrm{d} \omega \leq -\epsilon \lVert w \rVert^2
	\end{equation}
	is guaranteed.
\end{cor}
Note that, as shown in \cite{cavalcanti2020}, for multipliers of structure \(\bm{\Pi}\), condition \cref{eq:singular_valuecond} implies the following \acrlong{fdi}, which is known from earlier works on \gls{iqc} stability analysis. Letting \( \Pi_a =  \mathcal{C}(\Pi,\Pi_p) \),
\begin{equation} \label{eq:fdi}
	\begin{pmatrix}
		\mathcal{G} (i \omega) \\ I
	\end{pmatrix}^H \Pi_a (i \omega) \begin{pmatrix}
		\mathcal{G} (i \omega) \\ I
	\end{pmatrix} \leq -\epsilon I \quad \forall\; \omega \in \mathbb{R} \text{.}
\end{equation}
With this modification, the search for a feasible multiplier \(\Pi_a\) can be performed directly in the frequency domain using nonsmooth optimization techniques such as \cite{apkarian2006}, without having to resort to frequency gridding. The advantages over using the \gls{kyp} lemma to obtain an equivalent \gls{lmi} condition for \cref{eq:fdi} without gridding are manifold:
\begin{enumerate}[1)]
	\item No Lyapunov variables need to be introduced, which scale quadratically with system size and pose computational limitations for large systems.
	\item No \gls{sdp} solvers are required to solve any \glspl{lmi}. These tend to struggle with numerical issues when the involved matrices are badly conditioned and may falsely report an infeasible solution as feasible, posing risks if the result is not verified. Bad conditioning can be an outcome of transformations needed to render the problem convex, see e.g.\ \cite{schuette2022}.
	\item Non-convex problems (e.g.\ synthesis) can be directly solved. For the \gls{lmi} approach, iterative methods, e.g.\ \cite{veenman2014}, that cannot provide even local optimality certificates and produce often undesirable high-order controllers need to be used instead, increasing conservatism.
	\item Structural constraints on the controller design are easily incorporated.
	\item Poles of the basis functions for dynamic multipliers need not be selected a priori but are optimized by the optimization algorithm.
\end{enumerate}
Possible downsides of the frequency domain approach are addressed in \cref{sec:results}.

\section{Robust \(\mathcal{H}_2\)-Performance} \label{sec:h2_mult}
The issue of robust \(\mathcal{H}_2\)-performance in the sense of \( \sup_{\Delta \in \bm{\Delta}} \lVert \mathcal{G}_\Delta \rVert_2 \) is not trivial and has been discussed from various viewpoints in the literature. Noteworthy contributions include \cite{stoorvogel1993}, \cite{paganini1996}, \cite{feron1997} and \cite{sznaier2002}. A survey is given in \cite{paganini2000} and the topic is subject to ongoing research, see e.g.\ \cite{chamanbaz2020}. Despite the appealing nature of this performance metric from an engineering perspective, \(\mathcal{H}_\infty/\mathcal{H}_2\)-control, guaranteeing \(\mathcal{H}_2\)-performance only for the nominal plant and performing the worst-case analysis in terms of an \(\mathcal{H}_\infty\) specification, has emerged as a popular choice due to its simplicity, \cite{sznaier2002}. Complications in the mathematical treatment of \(\mathcal{H}_2\)-criteria arise particularly in the case of non-\gls{lti} uncertainties, where the classical definition of the \(\mathcal{H}_2\)-norm is not applicable and the interpretations of the \(\mathcal{H}_2\)-performance in terms of the impulse response of a system and its ability to reject white noise diverge, see \cite{paganini2000}. The case of white noise rejection is further complicated by the stochastic nature of the disturbance source and has strategically been adapted in \cite{paganini1996} by considering a worst-case analysis over a set of deterministic white signals \(W_{\eta,B} \subset \mathcal{L}_2^\bullet \) instead, resulting in the semi-norm \(\lVert \cdot \rVert_{W_{\eta,B}}\). These signals are characterized by having a flat spectrum with accuracy \(\eta\) up to bandwidth \(B\) and may taper off thereafter. Although it was shown that, quantitatively, the bound obtained from the white noise rejection approach can be greater than the bound obtained from the impulse response approach,
\begin{equation*}
	\lVert \mathcal{G}(\Delta) \rVert_{2,\mathrm{ir}} \leq \lVert \mathcal{G}(\Delta) \rVert_{2,\mathrm{wn}} \text{,}
\end{equation*}
it is important to keep in mind that these interpretations are no longer equivalent, i.e.\ neither choice is inherently superior. We adopt the white noise rejection interpretation going forward with
\begin{equation}
	\lVert \mathcal{G}_\Delta \rVert_{2,\mathrm{wn}} \coloneqq \lim_{\substack{\eta \to 0+ \\ B \to \infty}} \lVert \mathcal{G}_\Delta \rVert_{W_{\eta,B}}
\end{equation}
since its definition in terms of signals with particular spectral properties is more natural to the frequency domain approach pursued in this work. Furthermore, the computation of a bound on \( \lVert \mathcal{G}(\Delta) \rVert_{2,\mathrm{ir}} \) requires solving an \acrlong{are} and is therefore susceptible to similar computational scalability issues previously discussed for the \gls{kyp} lemma. Nevertheless, we note that, as shown in \cite{paganini2000}, the impulse response method weakly exploits the necessary causality of \(\Delta\), which will in general lead to less conservative bounds compared to the white noise rejection approach.

We now show how the result of \cite{paganini1999} may be transferred to the \gls{iqc} framework.
\begin{cor} \label{c:h2bound}
	For \( Y \in \mathcal{S}_\infty\) and of finite order, the multiplier
	\begin{equation} \label{eq:mult_h2}
		\Pi_p = \begin{pmatrix}
			I & 0     \\
			0 & -Y
		\end{pmatrix}
	\end{equation}
	in conjunction with Corollary \ref{c:iqc} guarantees robust worst-case \(\mathcal{H}_2\)-performance \(\gamma\) on the channel \(w \to z\) of interconnection \cref{eq:sys} with
	\begin{equation} \label{eq:rob_h2_bound}
		\lVert \mathcal{G}_\Delta \rVert_{2,\mathrm{wn}} \leq \frac{1}{2\pi} \int_{-\infty}^\infty \trace \left( Y(i \omega) \right) \,\mathrm{d} \omega < \gamma^2 \text{.}
	\end{equation}
\end{cor}
\begin{pf}
	First note that \(\Pi_p \in \bm{\Pi}\), so Corollary \ref{c:iqc} applies. Further, since \(Y(s) \in \mathcal{S}_\infty \) and of finite order, its frequency response \(Y(i \omega)\) is of bounded variation and there exists a monotonically decreasing function \(g \in \mathcal{L}_1 (\mathbb{R}_+) \) such that
	\begin{equation}
		0 \leq Y(i \omega) \leq g(\lvert \omega \rvert) I \text{,}
	\end{equation}
	where the first inequality holds as a consequence of \eqref{eq:fdi}.
	It is then straightforward to show that the sufficiency part of Theorem~6 in \cite{paganini1999} deriving the bound \cref{eq:rob_h2_bound} and its extensions to multivariable noise and nonlinear uncertainties also hold for \(Y(s = i \omega)\).
\end{pf}

\section{Implementation} \label{sec:implementation}
To synthesize a performance-optimized, robust and structured controller, we first formulate a parameterized optimization problem based on the theory in \cref{sec:iqc,sec:h2_mult} that can be solved with standard optimization algorithms. In this work, we rely on the nonsmooth optimization algorithm by \cite{apkarian2006} and its implementation in the \texttt{SYSTUNE} function from the MATLAB Control System Toolbox, \cite{matlab2021}.
\begin{prob} \label{prob:iqc}
	Given interconnection \cref{eq:sys} well-posed for all \(\Delta \in \bm{\Delta}\),
	\begin{align}
		\underset{\chi}{\text{minimize}} \quad & \frac{1}{2\pi} \int_{-\infty}^{\infty} \trace(Y(i \omega, \chi)) \,\mathrm{d} \omega \label{eq:traceint} \\
		\text{subject to} \quad &  \mathcal{G}(s,\kappa) \text{ stable,} \nonumber \\
		& \max_{\omega} \bar{\sigma} \left(\sect \left( - \mathcal{P}(s, \chi, \kappa) \right) \right) < 1 \text{,} \label{eq:sect_hard_cnst} \\
		& h(\chi) \leq 0 \text{, } l(\kappa) \leq 0 \text{.} \label{eq:mult_cnst}
	\end{align}
\end{prob}
where \( \mathcal{G}(s, \kappa) = \mathcal{F}_l (G(s), K(s,\kappa)) \) and
\begin{equation}
	\mathcal{P}(s, \chi, \kappa) =
	\begin{bmatrix}
		\phi_o (i \omega, \chi) \mathcal{G} (i \omega, \kappa) + \Psi (i \omega, \chi) & 0 \\
		\Sigma (i \omega, \chi) \mathcal{G} (i \omega, \kappa) & - \frac{1}{2} I
	\end{bmatrix} \text{.}
\end{equation}
Equations \cref{eq:traceint,eq:sect_hard_cnst} stem from Corollary~\ref{c:iqc} and Corollary~\ref{c:h2bound}, respectively. \(\chi\) and \(\kappa\) are real parameter vectors for the multiplier and controller variables, respectively, and we require that \(\Pi_a(s,\chi) \in \bm{\Pi}\) and \(K(s,\kappa) \in \varStructK\) are smooth in their respective parameters. The functions \(h(\cdot)\) and \(l(\cdot)\) serve as constraints on the parameter vectors that can, for example, be used to ensure \cref{eq:iqc} holds, i.e.\ \(\Pi\) a-priori satisfies the \gls{iqc} for a given class of uncertainty, or to further shape the set of admissible controllers \(\varStructK\) in addition to the structural dependence of \(K\) on \(\kappa\).

As of version 10.11 of the MATLAB Control System Toolbox, the \texttt{SYSTUNE} function does not support tuning goals of the form \cref{eq:traceint}. The results presented in \cref{sec:results} were produced with an experimental implementation where we extended the \texttt{SYSTUNE} code with the required functionality for a specific but nevertheless versatile parameterization for \(Y(s)\) adapted from \cite{jonsson1999}:
\begin{equation} \label{eq:Yparam}
	\begin{split}
		Y(s, \chi) &= \Psi_Y (s, \chi) + \Psi_Y (s, \chi)^H \\
		\Psi_Y (s, \chi) &= \sum_{i = 1}^{N} \frac{X_i s + X_i a_i + Z_i b_i}{s^2 + 2 a_i s + a_i^2 + b_i^2} \text{.}
	\end{split}
\end{equation}
Here \(X_i\) and \(Z_i\) are real matrices and \(a_i \in \mathbb{R}_+\) and \(b_i \in \mathbb{R}\) are scalars that determine the real and imaginary parts of the pole location of each summand, respectively. As such, as long as all \(a_i,b_i\) are distinct, \(\Psi_Y(s, \chi)\) is the most general way to parameterize an \(N\)\textsuperscript{th}-order transfer matrix in \(\mathcal{RH}_\infty\). The parameter vector \(\chi\) of \(Y(s, \chi)\) is just the concatenation of the vectorized values \(X_i, Z_i, a_i, b_i\).

In order to add a new tuning goal for \texttt{SYSTUNE}, an evaluation of the objective function \cref{eq:traceint} and its gradient in terms of its parameters has to be implemented. While numerical methods can be used here, the parameterization \cref{eq:Yparam} admits an elegant analytical solution.
\begin{lem}
	Given parameterization \cref{eq:Yparam} for \(Y(s,\chi)\), it holds that
	\begin{equation}
		\frac{1}{2\pi} \int_{-\infty}^{\infty} \trace (Y(i \omega, \chi)) \,\mathrm{d} \omega = \sum_{i = 1}^N \trace (X_i) \text{.}
	\end{equation}
\end{lem}
\begin{pf}
	Following the arguments in \cite{jonsson1999}, we find for each summand in \cref{eq:Yparam} that, for \(b_i = 0\),
	\begin{multline} \label{eq:lim_PsiY}
		\lim_{B \to \infty} \int_{-B}^{B} \frac{X_i s + X_i a_i + Z_i b_i}{s^2 + 2 a_i s + a_i^2 + b_i^2} \,\mathrm{d}\omega \\
		\begin{split}
			&= \lim_{B \to \infty} \int_{-B}^{B} \frac{X_i}{i\omega + a_i} \,\mathrm{d}\omega \\
			&= \lim_{B \to \infty} 2 X_i \arctan \left( \frac{B}{a_i} \right) = \pi X_i \text{.}
		\end{split}
	\end{multline}
	By analytic continuation of the function \( \int_{-B}^{B} \frac{1}{i\omega + z} \,\mathrm{d}\omega \) in the region \( \mathrm{Re} \, z > 0 \), \cref{eq:lim_PsiY} holds for all \( a_i > 0,\, b_i \in \mathbb{R} \). It follows that
	\begin{multline}
		\frac{1}{2\pi} \int_{-\infty}^{\infty} \trace(Y(i \omega, \chi)) \,\mathrm{d}\omega \\
		\begin{split}
			&= \frac{1}{\pi} \lim_{B \to \infty} \int_{-B}^{B} \trace(\Psi_Y (i \omega, \chi)) \,\mathrm{d}\omega \\
			&= \sum_{i=1}^{N} \trace(X_i) \text{.}
		\end{split}
	\end{multline}
\end{pf}
The gradient \( \tfrac{\mathrm{d}}{\mathrm{d}\chi} \tfrac{1}{2\pi} \int_{-\infty}^{\infty} \trace(Y(i \omega, \chi)) \mathrm{d} \omega \) is consequently easily computed as well - it is just one for every diagonal element of any \(X_i\) and zero otherwise. At this point, it may seem that in contrast to advantage~5, the other parameters including the pole and zero locations are not optimized, as they do not appear in the gradient. This however is not the case, as these parameters are still part of the hard constraint \cref{eq:sect_hard_cnst}.

Still, with respect to the conservatism of this method, we point out that unlike the frequency gridding approach used in \cite{paganini1997}, the finite-order parameterization \cref{eq:Yparam} of \(Y(s)\) cannot achieve tightness at all frequencies.

\section{Discussion \& Results} \label{sec:results}
To evaluate the practical use and limitations of the proposed method, we solve Problem~\ref{prob:iqc} for a simple artificial plant and a complex real-world system.

\subsection{\nth{4}-Order MISO System with Uncertain \\ Natural~Frequency and Measurement Time Delay} \label{sec:td_sys}
\begin{figure}
	\centering
	\includegraphics{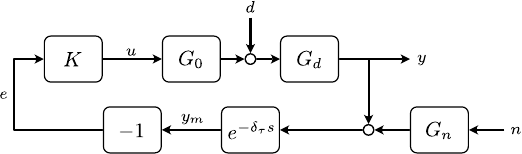}
	\caption{Demo system with parametric and measurement time delay uncertainty.}
	\label{fig:ex2_blockdiagram}
\end{figure}
This example, inspired by \cite{veenman2014} and depicted in \cref{fig:ex2_blockdiagram}, studies the robust control of a plant with arbitrarily fast time-varying natural frequency given by
\begin{equation}
	G_0 (s, \delta) = \frac{4}{s^2 + 0.1 (1 + 0.5 \delta) + (1 + 0.5 \delta)^2} \text{,}
\end{equation}
where \( \delta \in \bm{\Delta}_\mathrm{ltv,re,dr} \) with \(\lvert \delta \rvert \leq 1\) and \( \lvert \dot{\delta} \rvert \) unbounded. It is followed by an output disturbance modeled by
\begin{equation}
	G_d (s) = \frac{10}{s + 0.1} \text{.}
\end{equation}
The output measurement delay \(\delta_\tau \in \left[0, 0.025\right]\) is considered constant. As such, this uncertainty is of type \( \bm{\Delta}_\mathrm{ltv,rb,td} \) and we use the corresponding multiplier from the library of \cite{cavalcanti2020}. Instead of the original mixed-sensitivity problem, we adapt the example to an output regulation problem, i.e.\ we are interested in keeping the output at zero under the influence of white noise at the disturbance and measurement noise inputs. The latter was added in place of the reference input and is filtered by
\begin{equation}
	G_n(s) = \frac{s}{s + 10}
\end{equation}
to achieve the typical high-pass characteristic of measurement noise. As such the system caters nicely to the robust \(\mathcal{H}_2\)-performance problem. After \gls{lfr} decomposition into the form of \cref{fig:lfr}, the uncertainty \(\Delta\) is composed as
\begin{equation}
	\Delta = \begin{pmatrix}
		\delta I_3 & 0 \\
		0 & \delta_\tau
	\end{pmatrix} \text{.}
\end{equation}
The uncertainty \( \delta I_3 \in \bm{\Delta}_\mathrm{ltv,re,dr} \) is captured by the constant multiplier
\begin{equation}
	\Pi (s) = \begin{pmatrix}
		D^H D & W^H \\
		W & -D^H D
	\end{pmatrix} \text{,}
\end{equation}
where \(D \in \mathbb{R}^{3 \times 3} \) and
\begin{equation}
	W = \begin{pmatrix}
		0 & W_{12} & W_{13} \\
		- W_{12} & 0 & W_{23} \\
		- W_{13} & - W_{23} & 0
	\end{pmatrix} \in \mathbb{R}^{3 \times 3} \text{,}
\end{equation}
the particular structure of which is enforced with constraint \cref{eq:mult_cnst}.

\begin{figure}
	\centering
	\includegraphics{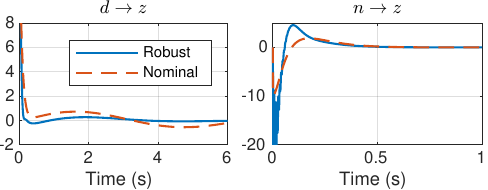}
	\caption{Impulse responses for the system in \cref{fig:ex2_blockdiagram} for a nominal and a robust controller of third order.}
	\label{fig:impresp}
\end{figure}
\cref{fig:impresp} shows the impulse responses of a third-order controller synthesized using \texttt{SYSTUNE} for the nominal plant without measurement delay and of a third-order robust controller synthesized using Problem~\ref{prob:iqc} when applied to a Padé-approximation of the system for \(\delta = -1\) and \(\delta_\tau = 0.025\). The nominal controller achieves a \(\mathcal{H}_2\)-norm of \num{4.58}, whereas the robust controller achieves \num{4.16}, an improvement of \SI{9}{\percent}. This is in fact the worst performance for both controllers that was found over a grid scan of \(\delta\) and on average, the nominal controller actually performs better by \SI{17}{\percent}. This however is purely empirical and no hard guarantee can be provided, whereas the robust controller provides a guaranteed worst-case bound of \num{6.85}.

\subsection{Decentralized Control for an Optical Synchronization System with Uncertain Dampings}
\begin{figure*}
	\centering
	\includegraphics{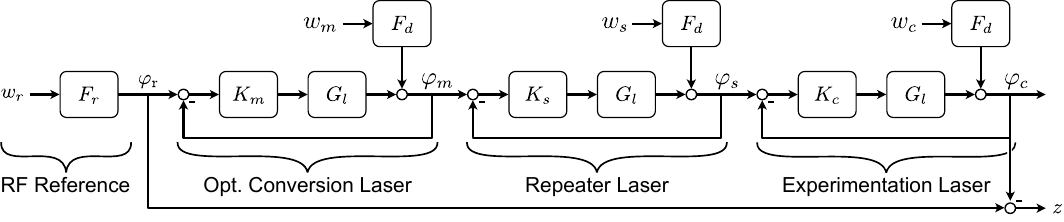}
	\caption{Block diagram of the \acrshort{lbsync} system.}
	\label{fig:lbsync}
\end{figure*}
Finally, the full flexibility of this controller synthesis method is demonstrated using a real-world control problem for a complex decentralized system. The \gls{lbsync} system, \cite{schulz2015}, developed at \gls{desy}, is used to transport an ultra-stable timing reference signal across a large-scale facility from a common root via intermediate oscillators to multiple end stations, creating a network of \glspl{pll} with local controllers. Using mode-locked lasers as tunable optical oscillators allows high-bandwidth locking to optical or \gls{rf} reference signals and the bidirectional property of optical fibers allows for an autocorrelation scheme, with which the propagation time on the transmission line can be stabilized against environmental disturbances. With these techniques, a timing synchronization of sub \SI{10}{\fs} \gls{rms} can be achieved over multiple kilometers, which has been demonstrated in the case of the \gls{euxfel} photon-science facility in Hamburg, Germany, \cite{schulz2019}.

The associated control problem was classified as a decentralized, fixed-order, output-feedback problem in \cite{schuette2021} and extended with robustness guarantees in \cite{schuette2022}. To compare the \gls{lmi} synthesis method used in the latter work, which is based on convexification and transformation of the original problem which can then be solved with a single \gls{sdp}, we apply the frequency domain method of \cref{sec:implementation} to the same model of the \gls{lbsync} system that was used before. In this baseband model, a reference signal produced by an \gls{rf} oscillator is transferred to the optical domain using a first laser, repeated using a second laser for increased transmission distance, and finally used to synchronize an experimentation laser. As the phase-noise of each oscillator is independent, this yields a four-input, single-output system of \nth{24}-order, where the output is given as the phase error between the reference and the experimentation laser. Further details on the involved models are given in \cite{schuette2022}. The uncertainty is of class \(\bm{\Delta}_\mathrm{ltv,re,dr}\) and comprises independent uncertain dampings in the piezoelectric actuators used to tune each of the lasers. We note that the uncertainty classes that can be treated with the full-block multiplier approach used in \cite{schuette2022} is limited compared to the more general \gls{iqc} framework incorporated in this work, e.g.\ the example of \cref{sec:td_sys} is not covered because of the uncertain time delay.

\begin{table}
	\begin{threeparttable}
		\caption{Comparison of the \gls{lmi} and frequency domain synthesis solutions.} \label{tab:results}
		\begin{tabularx}{\linewidth}{Xrrr}
			\toprule
			Test & \gls{lmi} & Freq.-Domain & Rel.\ Impr.\\
			\midrule
			Robust \(\mathcal{H}_2\)-perf.\ bound & 29.154 & 9.908 & \SI{66.0}{\percent} \\
			Average \(\mathcal{H}_2\)-perf.\tnote{a} & \num{7.792} & \num{7.185} & \SI{7.8}{\percent} \\
			Worst \(\mathcal{H}_2\)-perf.\tnote{a} & \num{8.031} & \num{7.324} & \SI{8.8}{\percent} \\
			Gap & 21.123 & 2.584 & \SI{87.8}{\percent} \\
			Relative gap & \SI{263}{\percent} & \SI{35.3}{\percent} & \SI{86.6}{\percent} \\
			Comp.\ time per iter.\tnote{b} & \SI{2.48}{\second} & \SI{33.21}{\second} & \\
			\bottomrule
		\end{tabularx}
		\begin{tablenotes}
				\item[a] \num{10000} uncertainty samples
				\item[b] \SI{3.6}{\giga\hertz} six-core processor
		\end{tablenotes}
	\end{threeparttable}
\end{table}
The goal is to synthesize feedback gains for the individual \glspl{pll} locking the laser oscillators to their respective references that minimizes the robust \(\mathcal{H}_2\)-norm of the closed loop system. For dampings in the range \(\delta_i \in \left[ 0.1, 0.3 \right] \), results and statistics are collected in \cref{tab:results}. Besides an improvement of the worst-case nominal \gls{lti} \(\mathcal{H}_2\)-performance \( \sup_{\Delta \in \bm{\Delta}} \lVert \mathcal{G}_\Delta \rVert_2 \) of almost \SI{10}{\percent}, a clear advantage of the frequency domain method appears to be the significantly reduced gap between the provided bound on the robust performance and its statistically estimated actual quantity of over \SI{85}{\percent}. It is important to note at this point, with respect to the discussion in \cref{sec:h2_mult}, the bound provided by the \gls{lmi} approach is of the impulse-response type and therefore not strictly equivalent to the worst-case white noise rejection analysis result of the frequency domain approach. We also recall that the \gls{lmi} bound is computed only for a convexified version of the original problem, explaining the increased conservatism compared to the solution obtained from the nonsmooth optimization algorithm, which provides at least a local optimality certificate. An apparent disadvantage of the frequency domain method is its significantly increased computation time. Multiple comments are in order regarding this observation.
\begin{enumerate}[1)]
	\item To get a fair comparison, both computations were executed on the same machine, but we also had success running the modified \texttt{SYSTUNE} code parallelized on a \acrlong{hpc} cluster, greatly reducing the computation time.
	\item To reduce the number of random initial iterates, a hot-start technique may be considered, where a nominal controller is first synthesized and then randomly perturbed versions are used as initial points for the robust synthesis problem.
	\item The order of the discussed problem is still relatively low. Scaling issues of the \gls{lmi} approach are only expected to be a significant issue until the problem size exceeds a few hundred states, at which point the frequency domain method is expected to get the upper hand in terms of computational complexity.
\end{enumerate}

\section{Conclusion} \label{sec:conclusion}
In this paper, we demonstrated the successful application of a frequency domain \(\mathcal{H}_2\)-performance bound to robust controller synthesis by formulating a corresponding \gls{iqc} problem which is then solved using nonsmooth optimization. We proposed a suitable parameterization for the bounding multiplier which allows the bound and its gradient to be efficiently evaluated and used this to extend the applicability of MATLAB's \texttt{SYSTUNE} function to robust \(\mathcal{H}_2\) synthesis for a wide range of uncertainties. We evaluated this method using a modified literature example and verified the generated solutions using industry-standard robust analysis methods. We found that, using a moderate number of random initial parameters, robust controllers that outperform nominal solutions for certain uncertainty realizations could be synthesized. At the same time, we demonstrated the flexibility of the method in dealing with uncertainties from a wide range of classes. Finally, we applied the method to a complex real-world distributed system and found that, compared to a previously used \gls{lmi} synthesis method, the conservatism induced by using finite-order parameterizations for the \(\mathcal{H}_2\)-performance multiplier was significantly lower than the conservatism stemming from a convexification of the optimization problem for \gls{lmi}-based synthesis.

\bibliography{iqc_rob_h2_dist_freq}

\end{document}